\def\eeq{\end{equation}}
\def\beq{\begin{equation}}
\def\bea{\begin{eqnarray}}
\def\eea{\end{eqnarray}}

\documentstyle[aps]{revtex}

\begin{document}

\title{Asymmetric Unimodal Maps: Some Results from $q$-generalized 
Bit Cumulants}

\author{U\v{g}ur 
T{\i}rnakl{\i}\thanks{e-mail: tirnakli@sci.ege.edu.tr}}

\address{Centro Brasileiro de Pesquisas F\'\i sicas, Rua Xavier Sigaud 
150, 22290-180 
Rio de Janeiro - RJ, Brazil\\
and \\
 Department of Physics, Faculty of Science, Ege University,
35100 Izmir-Turkey\\}

\maketitle

\begin{abstract}
In this study, using $q$-generalized bit cumulants ($q$ is the 
nonextensivity parameter of the recently introduced Tsallis statistics), 
we investigate the asymmetric unimodal maps $x_{t+1} = 1-a|x_t|^{z_i}$ 
($i=1,2$ correspond to $x_t>0$ and $x_t<0$ respectively, $z_i >1$, 
$0<a\leq 2$, $t=0,1,2,...$). The study of the $q$-generalized second 
cumulant of these maps allows us to determine, for the first time in 
the literature, the dependence of the inflexion parameter pairs 
($z_1,z_2$) to the nonextensivity parameter $q$. This behaviour is 
found to be very similar to that of the logistic-like maps 
($z_1=z_2=z$) reported recently by Costa et al. 
[Phys. Rev. E 56 (1997) 245].\\

\noindent
{\it PACS Number(s): 05.45.-a, 05.20.-y, 05.70.Ce}

\end{abstract}


\vspace{1.5cm}

In recent years, sensitivity to initial conditions of nonlinear
dynamical systems has started to be studied with an increasing interest.
Among them, dissipative systems of low-dimensional maps 
\cite{TPZ,costa,lyra,circle}, self-organized criticality \cite{SOC}, 
symbolic sequences \cite{grigo} and conservative systems of long-ranged 
many-body Hamiltonians \cite{celia,latora}, low-dimensional maps 
\cite{latora2} can be enumerated. We focus here on the one-dimensional 
dissipative maps. As it has already been well studied in the literature, 
to investigate the sensitivity to initial conditions of any kind of
one-dimensional maps on the onset to chaos, it is possible to introduce 

\beq
\xi (t) = \lim_{\Delta x(t)\rightarrow 0} 
\frac{\Delta x(t)}{\Delta x(0)} \;\;\; ,
\eeq 
(where $\Delta x(0)$ and $\Delta x(t)$ are discrepancies of the initial
conditions at times $0$ and $t$) which, if the Lyapunov exponent 
$\lambda_1 \neq 0$, satisfies the differential equation 
$d\xi /dt = \lambda_1 \xi$, thus

\beq
\xi (t) = e^{\lambda_1 t}\;\;\; ,
\eeq
whereas, if $\lambda_1 =0$, it satisfies $d\xi /dt=\lambda_q\xi^q$, thus

\beq
\xi (t) = \left[1+(1-q)\lambda_q t\right]^{\frac{1}{1-q}}\;\;\;\;\;
(q\in {\cal R})
\eeq
which recovers the standard case (namely Eq.(2)) for $q=1$. Here, $q$ is
the nonextensivity parameter of recently 
introduced Tsallis statistics \cite{tsallis} and for $q\neq 1$ case, it
is evident that Eq.(3) yields a {\it power-law} sensitivity to initial
conditions. At the onset of chaos, $\xi (t)$ presents strong
fluctuations reflecting the fractal-like structure of the critical
attractor and Eq.(3) delimits the power-law growth of the upper bounds
of $\xi (t)$. These upper bounds ($\xi \propto t^{\frac{1}{1-q}}$) 
allow us to calculate the values of $q$ for the map under consideration. 
This method of finding numerical values of $q$ has been successfully 
used for the logistic map \cite{TPZ}, a family of logistic-like maps
\cite{costa}, the circle map \cite{lyra} and a family of circular-like
maps \cite{circle}. Beside this method, another one has been developed
by Lyra and Tsallis \cite{lyra} by looking at the geometrical aspects of
dynamical attractors at the treshold to chaos. Using the multifractal
singularity spectrum $f(\alpha)$ \cite{falpha}, they proposed the
scaling relation 

\beq
\frac{1}{1-q} = \frac{1}{\alpha_{min}} -
\frac{1}{\alpha_{max}}
\eeq
where $\alpha_{max}$ ($\alpha_{min}$) is the most rarefied 
(concentrated) region of the multifractal singularity spectrum of the
attractor. This relation presents the second method of calculating the
$q$ values once the scaling properties of the dynamical attractor are
known. It has already been shown that for all above-mentioned 
one-dimensional dissipative maps, the values of $q$ index calculated 
within these two different methods are the 
same (within a good precision) for any given value of the inflexion
parameter $z$ of the used map. In addition to this, another important
information which comes from all these works is the determination of the
behaviour of the $q$ index as a function of the inflexion parameter
$z$. As it can be seen from Fig.2 of \cite{costa} and from the Table of
\cite{circle}, it seems that when $z\rightarrow \infty$ , $q$
approaches 1.

The purpose of this paper is to show numerically whether such kind of
behaviour is verified for asymmetric unimodal maps (AUMs)

\begin{eqnarray}
x_{t+1} = \left\{ \begin{array}{ll}
                  1- a|x_t|^{z_1}  & \mbox{if $x_t>0$} \\
                  1- a|x_t|^{z_2}  & \mbox{if $x_t<0$}
             \end{array} \right.
\end{eqnarray}
where $z_{1,2} >1$, $0<a\leq 2$ and $t=0,1,2,...$. Unfortunately, up to
now, for the AUMs, neither the first nor the second method can yield 
satisfactory results for the prediction of $q$ values of any inflexion 
parameter pair ($z_1,z_2$) \cite{unpub}. The problem of the first 
method might be originated from the prediction of the critical $a_c$ 
values at the chaos treshold with enough precision, whereas the problem 
of the second method might be related to the numerical procedure used 
to estimate the $f(\alpha)$ curve, whose most rarefied region 
($\alpha_{max}$) is usually poorly sampled. 
Another fact which yields this failure might be the nonuniversal 
behaviour of AUMs reported in \cite{ma,tsa} 
(for example, it is shown that AUMs fail to exhibit the metric
universality of Feigenbaum). Since the values of $q$ are not available
for AUMs, clearly it is not possible to see the behaviour of $q$ 
as a function of ($z_1,z_2$) pairs by using the above-mentioned two
methods. In this paper, in order to see this behaviour, {\it without}
finding the precise values of $q$ for ($z_1,z_2$) pairs, we use another
technique based on the very recent generalization of bit cumulants for
chaotic systems within Tsallis statistics \cite{ramanrai,rairaman}. 
To explain this technique, let us recall the main results of 
\cite{ramanrai,rairaman}. The generalized second cumulant (or heat
capacity) is given by 

\beq
C_2^{(q)} = \frac{q}{\left(q-1\right)^2}
\left[\left<\rho^{2q-1}\right> - \left<\rho^q\right>^2\right]
\eeq
where $\rho$ is the natural invariant density \cite{beck}. 
As $q\rightarrow 1$, we have 

\beq
 C_2^{(1)} = 
\left<\left(\ln\rho\right)^2\right> - \left<\ln\rho\right>^2
\eeq
which is equivalent to the standard definition of the second cumulant 
\cite{beck,schogl}. It is shown \cite{ramanrai,rairaman} that there is 
a kind of scaling between $C_2^{(1)}$ and $C_2^{(q)}$ which is evident
from a $C_2^{(1)}$ vs $C_2^{(q)}$ plot, where it is easily seen that
most of the points can be fitted to a straight line. The slope of this
line gives the scaling factor between $C_2^{(1)}$ and $C_2^{(q)}$ and
also provides us a tool to fulfill our aim in this paper. 
In Fig.1, we illustrate this slope plotting $C_2^{(1)}$ 
vs $C_2^{(q)}$ curve for a representative ($z_1,z_2$) pair. 
It is evident that most of the data points fall onto a straight line 
which yields a well-defined slope 
(in fact, we encountered that a few points deviate from this straight
line especially when the values of the inflexion parameters start to be 
very different from the standard case (namely, $z_1=z_2$), but since
such data points are very few, we prefer to calculate the slope using a
linear regression --with an error of $\pm 0.01$ for each estimation of
the slope-- to data points which fall onto this straight line). 
As it is 
evident from Fig.2, when $q\rightarrow 1$, naturally  
$C_2^{(q)}\rightarrow C_2^{(1)}$ and thus the slope also tends to 
unity. Therefore, now we have another parameter (which behaves exactly 
the same as $q$) from where we can check the behaviour of the $q$ 
index as a function of ($z_1,z_2$) pairs 
{\it without knowing the exact values of $q$}! 
Calculating the above-mentioned slope for various values of ($z_1,z_2$) 
pairs, we are able to estimate the behaviour of $q$ index as a 
function of inflexion parameters without finding any $q$ value. 
Fig.3 represents this behaviour for ($2,z_2$),($z_1,2$),($3,z_2$) 
and ($z_1,3$) cases. It seems from the figure that 
as $z_2 - z_1 \rightarrow \pm\infty$, 
the above-mentioned slope (and thus, $q$ index) 
{\it approaches to unity}! This tendency is exactly the same as that 
observed for a family of logistic-like maps \cite{costa} and a family 
of circular-like maps \cite{circle}, namely, 
as $z\rightarrow\infty$, $q$ approaches 1.

Summing up, for the first time (to the best of our knowledge), we manage
to study the asymmetric unimodal maps in such a way that it became
possible to relate it with the nonextensivity parameter $q$ of Tsallis
statistics. We were able to show that the dependence of the inflexion
parameter pairs ($z_1,z_2$) of these maps to the $q$ index is very
similar to that observed for other one-dimensional maps reported so far. 
Although the determination of the exact values of $q$ index for 
($z_1,z_2$) pairs is still lacking, we hope that present work would be 
considered as a first attempt on this line and will accelarate other
studies which are, no doubt, highly welcome.

\section*{Acknowledgments}
I acknowledge the partial support of BAYG-C program 
of TUBITAK (Turkish agency) and CNPq (Brazilian agency) as well as 
the support from Ege University Research Fund under the project 
number 98FEN025.



\vspace{1.5cm}

{\bf Figure Captions}

\vspace{1.5cm}

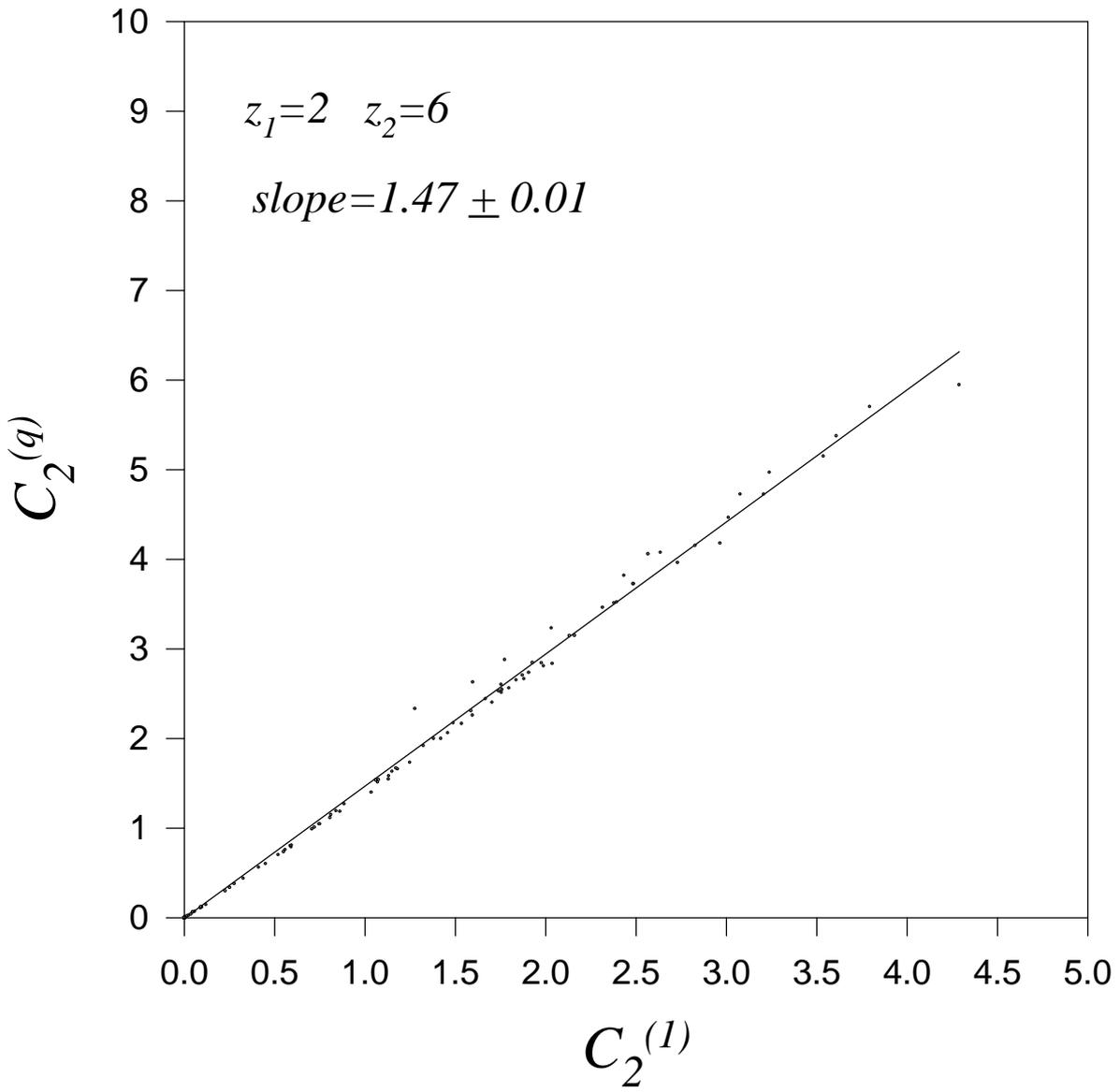
Figure 1 : The scaling between the standard second cumulant $C_2^{(1)}$
and the generalized second cumulant $C_2^{(q)}$ for a representative
value of ($z_1,z_2$) pairs.

\vspace{1cm}

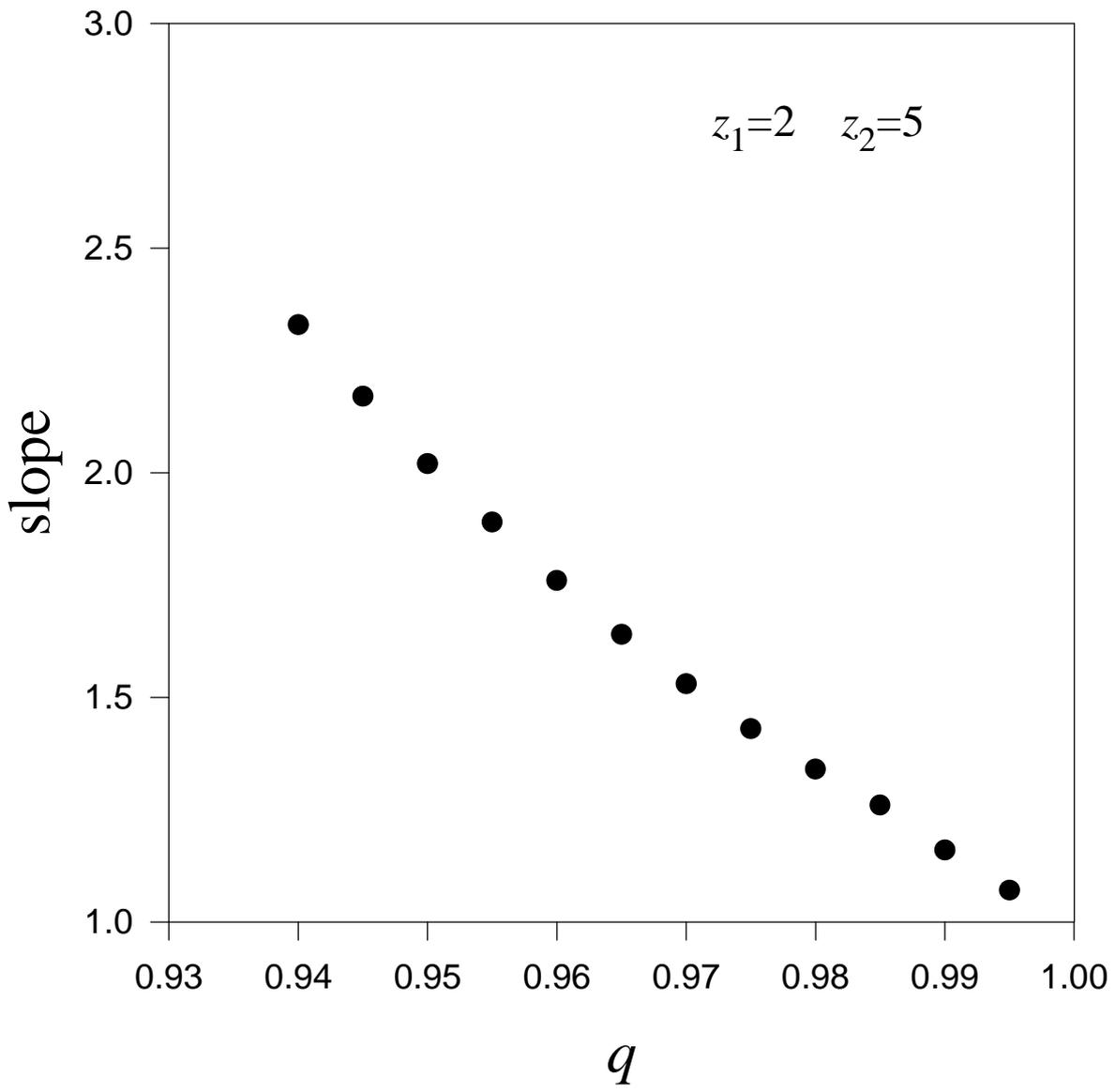
Figure 2 : The behaviour of the slope as a function of $q$ index for a
representative value of ($z_1,z_2$) pairs.

\vspace{1cm}

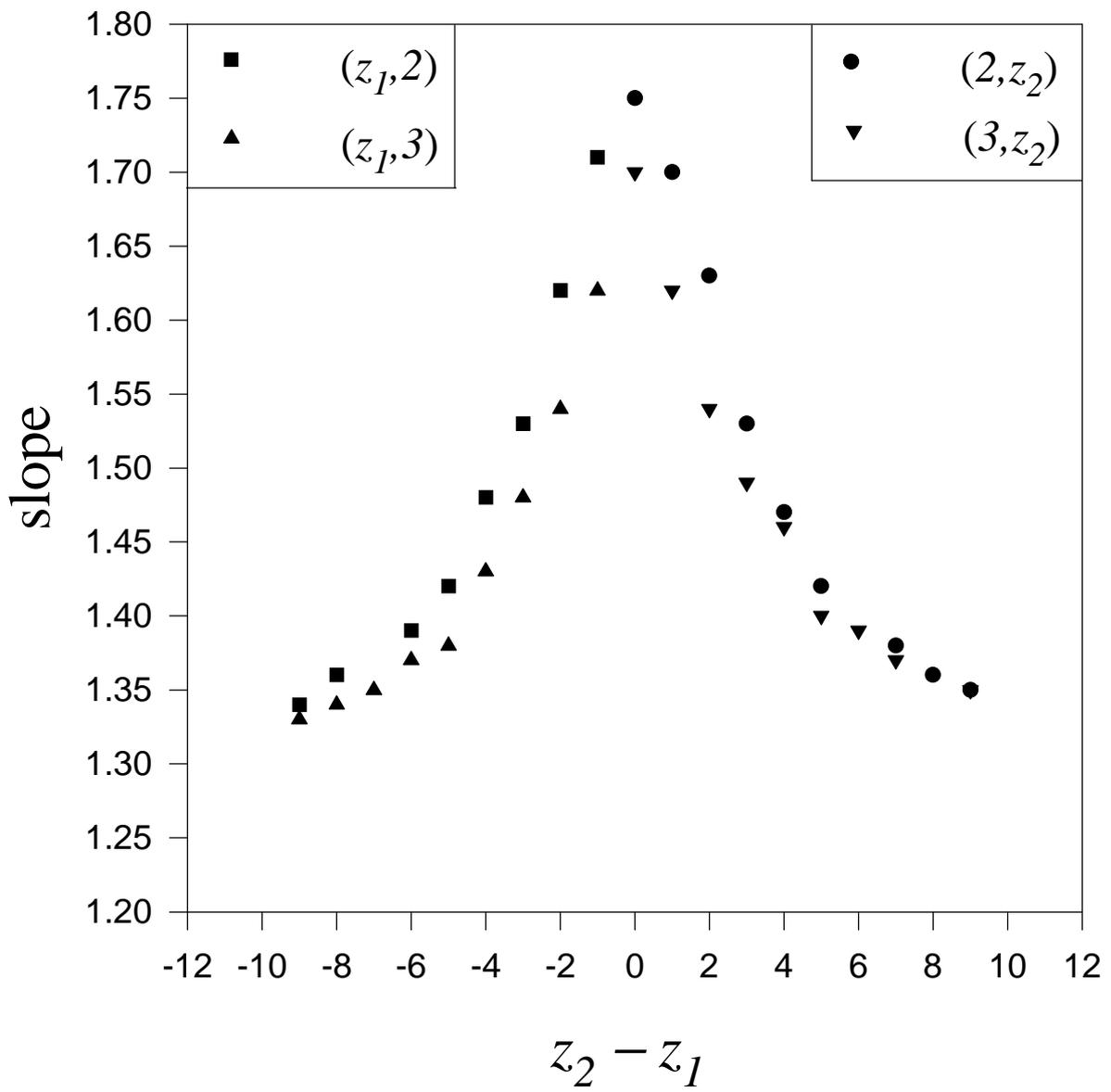
Figure 3 : The behaviour of the slope as a function of $z_2 - z_1$
values for a family of four different ($z_1,z_2$) pairs.

\vspace{1cm}

\end{document}